\documentclass[prb,twocolumn,showpacs,superscriptaddress]{revtex4-1}

\usepackage{graphicx}  
\usepackage{dcolumn}   
\usepackage{bm}        
\usepackage{amssymb}   
\usepackage{amsmath}

\usepackage{color}
\usepackage[usenames,dvipsnames]{xcolor}

\usepackage{epstopdf} 

\usepackage{natbib}
\usepackage[final=true]{hyperref}

\begin{document}
\title{Phonon drag thermopower in graphene in equipartition regime}

\author{S.~V.~Koniakhin}
\email{kon@mail.ioffe.ru}
\affiliation{Ioffe Physical-Technical Institute of the Russian Academy of Sciences, 194021 St.~Petersburg, Russia}
\affiliation{St. Petersburg Academic University - Nanotechnology Research and Education Centre of the Russian Academy of Sciences, 194021 St. Petersburg, Russia}

\author{E.~D.~Eidelman}
\affiliation{Ioffe Physical-Technical Institute of the Russian Academy of Sciences, 194021 St.~Petersburg, Russia}
\affiliation{St. Petersburg Academic University - Nanotechnology Research and Education Centre of the Russian Academy of Sciences, 194021 St. Petersburg, Russia}
\affiliation{St.~Petersburg Chemical Pharmaceutical Academy, 197022 St.~Petersburg, Russia}

\date{\today}

\begin{abstract}
This letter calculates the contribution of electron-phonon interaction to thermoelectric effects in graphene. One considers the case of free standing graphene taking into account interaction with intrinsic acoustic phonons. The temperatures considered range from liquid nitrogen to the room level. It has turned out that the contribution to thermoelectromotive force due to electron drag by phonons is determined by the Fermi energy in the sample and phonon relaxation time. The explicit temperature dependence of the contribution to thermoelectromotive force deriving from  by phonons is weak in contrast to that due to diffusion, which is directly proportional to temperature. The dependence obtained suggests that at the temperatures considered, a high carrier concentration and for samples with a high thermal conductivity, the phonon drag contribution can become dominant. Thus a theoretical limit  has been established for a possible increase of the thermoelectromotive force through electron drag by the intrinsic phonons of graphene.

\end{abstract}


\maketitle

The transport properties of graphene have been attracting considerable interest in recent years \cite{RevModPhys.81.109,RevModPhys.83.407,BalReview,NikaBalReview}. A study was performed of the electrical \cite{PhysRevB.82.195403,PhysRevLett.105.256805} and thermal conductivity \cite{6235226,PhysRevB.79.155413,Seol09042010,doi:10.1021/nl102923q}, thermomagnetic effects  \cite{PhysRevLett.107.016601} and  electrical phenomena generated by irradiation \cite{PhysRevB.86.115301,PhysRevB.84.125429}.

Particularly promising for the field of applications appear to be experimental \cite{PhysRevB.80.081413,PhysRevLett.102.166808,PhysRevLett.102.096807,PhysRevB.83.113403,Gabor04112011} and theoretical \cite{PhysRevB.79.075417,PhysRevB.82.155410,PhysRevB.80.235415,vaidya:093711,PSSB:PSSB201248302,PhysRevB.86.075411} studies of the thermoelectric properties of graphene. The high electrical conductivity combined with a significant Seebeck coefficient (thermoelectromotive force or thermopower), provide a basis for expecting graphene to reveal a fairly high thermoelectric figure of merit. Scattering from phonons is essential for evolvement of electron transport in graphene \cite{PhysRevB.79.235406,PhysRevB.81.121412,PhysRevB.82.195403,PhysRevB.65.235412,PhysRevB.72.235408,PhysRevLett.105.256805}. Thus, in graphene, as in other $sp^2$ carbon materials \cite{EidVul,PhysRevB.21.2462,PhysRevB.28.2157,PhysRevB.34.4298,PhysRevB.66.205405}, one can envisage amplification of thermoelectric effects as a result of the phonon drag effect discovered in 1946 by L.E. Gurevich  \cite{Gurevich46}.

We calculate the electric current generated by phonon drag in the presence of a temperature gradient in a graphene sample, obtain the corresponding contribution to the Seebeck coefficient and compare it with the contribution of diffusion to thermopower.

The conditions chosen for the calculations are the most favorable for experimental observation of the effect of electron drag by phonons. The assumptions accepted for the sample and external conditions in the system are as follows. We consider electrons near the Dirac point having an idealized linear dispersion law $E_{\bm k} = \hbar v_F k$ (the Fermi velocity $v_F\approx10^{8}\mathrm{cm s}^{-1}$). The Fermi level $\varepsilon_F$ in the system is assumed to be much higher than the temperature, opening a way to application of the degenerate electron gas approximation, which implies monopolar conduction. The condition of the onset of degeneracy defines the upper limit of temperature below which this theory is applicable. The absolute value of the electron wave vector $k$ near the Fermi level will be denoted in what follows by $k_F$. We are going to consider a free-standing graphene monolayer and take into account the contribution of the intrinsic acoustic in-plane (LA and TA) phonons only, which correlates with previous studies \cite{PhysRevB.79.235406} of the mechanisms underlying electron relaxation in graphene. The dispersion law of such lattice vibrations is given by $\hbar\omega_{ph}(q)=\hbar q v_s$, where sound velocity is $v_s\sim 2\cdot10^{6}\mathrm{cm~s}^{-1}$.

On the other hand, it is assumed that the temperature is high enough to maintain an equipartition (EP) regime of electron-phonon interaction, in which the numbers of occupation of the phonon modes contributing to the drag effect are larger than 1. The phonons interacting with electrons have a wave vector on the order of $k_F$ and energy $\hbar v_s k_F$.

In the case of high temperatures, the equilibrium distribution function for phonon mode occupation numbers transforms into  $N^{(0)}_{ph}(\bm q)=T/\hbar \omega_{ph}(\bm q)$ for both phonon emission and absorption processes. This defines the lower temperature limit of applicability of the theory developed. Only the contribution from acoustic phonons is considered, and, hence, the occupation numbers of optical phonons must be small. This condition specifies the second upper limit in temperature for the theory developed. Fig.\,1 demonstrates all characteristic energies, wave vectors and velocities related to the problem being analyzed. We readily see that in actual fact the above conditions are satisfied by the very broad temperature range, including room temperature. Finally, we neglect the reverse influence of the electron subsystem of the crystal on its phonon subsystem.

The Fermi energy of graphene $\varepsilon_F$ is known to be related \cite{RevModPhys.83.407} with carrier concentration $n_s$ as $\varepsilon_F=\hbar v_F \sqrt{\pi n_s}$. In transistor-based electronic devices the concentration is controlled by properly varying the gate voltage and is determined by the capacity of the structure. These parameters are measured in an experiment. Through them can be expressed all the results obtained. Doping is another way to control the Fermi level in graphene \cite{PhysRevLett.101.026803}.

\begin{figure}
\includegraphics[width=1.0\linewidth]{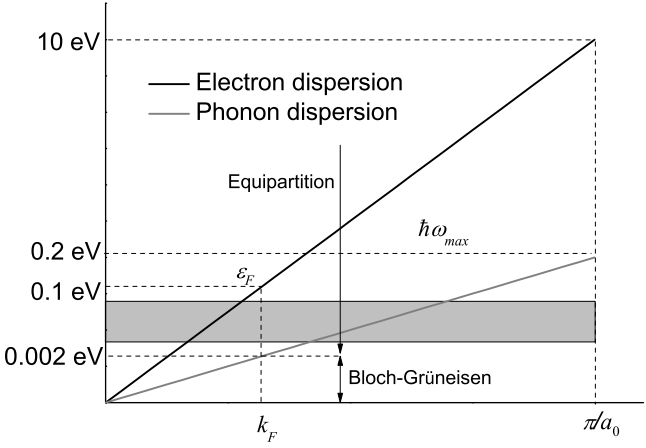}
\caption{Schematic of the energy scales essential for the problem under study. Dispersion of electrons (black line) and dispersion of acoustic phonons (grey line) are shown on an arbitrary scale. The $\Gamma$ point in the Brillouin zone for phonons and the K point in the Brillouin zone for electrons are superposed at the origin. Characteristic energies, such as the zone width $\sim10\mathrm{eV}$, maximum phonon energy $\hbar \omega_{max}\sim0.2\mathrm{eV}$, Fermi energy $\varepsilon_F$ and energy of acoustic phonons with wave vector to $k_F$ are specified. Also presented are numerical estimates of these quantities. The gray band confines the temperature range of applicability of the theory developed. The value $0.1\mathrm{eV}$ for Fermi energy was chosen because it provides the degeneracy of electron gas a room temperature ($0.026\mathrm{eV}$). The energy of acoustic phonons with comparable wave vectors is $0.002\mathrm{eV}$, which corresponds to 23 K. It separates equipartition and Bloch-Gr\"{u}neisen temperature regimes.}

\label{fig1}

\end{figure}

A temperature gradient in the sample produces a directed phonon flow (the so-called "phonon wind"). The behavior of the phonon system in the presence of a temperature gradient can be described in the relaxation time approximation with the Boltzmann kinetic equation. The kinetic equation describing the correction  $N^{(1)}_{ph}(\bm q) = N_{ph}(\bm q) - N^{(0)}_{ph}(\bm q)$ to the equilibrium distribution function for phonon mode occupation numbers in the presence of temperature gradient reads as
\begin{equation} \label{kinetic_phon}
v_s \frac{\bm q}{q} \frac{\partial N^{(0)}_{ph}(\bm q)}{\partial T} k_B \nabla T = -\frac{N^{(1)}_{ph}(\bm q)}{\tau_{ph}(\bm q)}\,,
\end{equation}

and, hence, the correction itself can be written as

\begin{equation} \label{correction_phon}
N^{(1)}_{ph}(\bm q) =- \frac{\tau_{ph}(\bm q)k_B}{\hbar q^2}(\bm q \nabla T)\,,
\end{equation}
where $\tau_{ph}(\bm q)$ is the relaxation time of phonon with wave vector $\bm{q}$.

Here the relaxation time should be taken for the in plane acoustic phonons with a wave vector magnitude from 0 to $2k_F$. Within the temperature range treated in this letter, phonon scattering from point defects and sample edges are dominant mechanisms \cite{6235226,PhysRevB.79.155413,PhysRevB.79.075417,PhysRevB.82.155410}. For a sample 2 $\mu$m in size and with the same concentration of point defects as observed in Ref. \cite{6235226}, scattering from edges dominates over that from point defects. The relaxation time for edge scattering can be estimated as the ratio of the minimum linear size of the sample $L$ (de-facto mean free path) to the sound velocity $v_s$. For the corresponding time for a $2 \mathrm{\mu m }$ sample one obtains $100\,\mathrm{ps}$, and it does not depend on the magnitude of the phonon wave vector $q$ for in plane phonons, thus permitting us to write simply in what follows $\tau_{ph}$.

Thermal conductivity of graphene reaches the value of 5000\, $\mathrm{Wm}^{-1}\mathrm{K}^{-1}$ \cite{6235226,PhysRevB.79.155413,BalReview,NikaBalReview}. In a general case, the relaxation time of phonons that generates the drag effect can not be straightly connected with thermal conductivity of the sample because the phonons from whole Brillouin zone, the phonons from other branches and other mechanisms of phonon scattering (including the three-phonon Normal and Umklapp processes at the temperature higher than 300\,K) define total thermal conductivity of the sample.

However, the value of the relaxation time of the phonons generating the drag effect can be connected with low temperature ($T<30\,$K) thermal conductivity of the sample. It is known that the thermal conductivity of graphene at low temperatures is provided by ZA phonons and corresponding relaxation mechanism is scattering on the edges of the sample. The dispersion relation of ZA phonons is quadratic \cite{6235226}: $\omega_{ZA}(q)=K q^2$, where $K=3.13\cdot 10^{-3}\,\mathrm{cm^2 s^{-1}}$ \cite{PhysRevB.68.134305}. The relation (9) from \cite{6235226} after substitution the relaxation time $\tau_{ZA}(q)=L/v_{ZA}(q)$ and some algebra connects the phonon mean free path $L$ and thermal conductivity $\kappa$:

\begin{equation} \label{heat_conductivity}
\kappa = 0.35\frac{k_B^{5/2}T^{3/2}}{K^{1/2}h\hbar^{3/2}}L,
\end{equation}
where $h \sim 0.35\,\mathrm{nm}$ estimates the thickness of graphene layer \cite{PhysRevB.79.155413}. This expression gives approximately $900\,\mathrm{W m^{-1}K^{-1}}$ for the sample with $L=2 \mathrm{\mu m}$ at 50\,K.

The rate of electron transitions from a state with wave vector $\bm k$ into a state with wave vector $\bm k^{\prime}=\bm k + \bm q$ driven by scattering from a phonon with wave vector $\bm q$ (involving absorption or emission of one acoustic phonon) is described by the "Fermi golden rule":

\begin{multline} \label{Fermi_golden}
W_{\bm k \rightarrow \bm k + \bm q}^{em,abs} = \frac{2 \pi}{\hbar} \frac{N_{ph}(\bm q)}{v_F \omega_{ph}(\bm q)} |M_{e-ph}(\bm q)|^2\\
\times\frac{S}{N_u M} \delta \left( k^\prime - k \pm \frac{\omega_{ph} (\bm q)}{v_F} \right)\,,
\end{multline}
where $M_{e-ph}(\bm q) $ is the matrix element of electron scattering from a phonon specifying the magnitude of electron--phonon interaction, $M$ is the carbon atom mass, $S$ the sample area, and $N_u$ is the number of elementary cells in the sample. The combination of the three last quantities can be expressed through the surface density of graphene $\rho=6.5\cdot 10^{-8} \mathrm{g\,cm^{-2}}$ and the lattice constant $a_0=2.46\mathrm{\AA}$.

For long-wavelength acoustic phonons the modulus of the matrix element depends on the phonon wave vector and the scattering angle $\theta_{\bm k + \bm q, \bm k}$ as $M_{e-ph}(\bm q)=Dq \frac{1}{2} (1+\cos (\theta_{\bm k + \bm q, \bm k} ))$ ~\cite{PhysRevB.65.235412,PhysRevB.76.045430,PhysRevB.79.235406}. The term in brackets suppresses electron backscattering \cite{PhysRevB.79.235406}. In the case of completely isotropic electron scattering, $M_{e-ph}(\bm q)=Dq$, the phonon drag thermopower would increase fourfold. The quantity $D$ can be estimated as 16\,eV. Note that this value is very close to the combination $\hbar v_F 2 \pi/a_0$, which permits the estimation of the graphene electron bands.

Should a more accurate expression for matrix element  $M_{e-ph}$ as a function of the wave vectors of electrons and phonons be required, it can be readily derived through the deformation potential precisely calculated in Ref.~\cite{PhysRevB.72.235408}.

In the first order of perturbation theory the contribution to the drag comes from 4 types of electron transitions, which correspond to the coming and leaving of electrons into the state with wave vector $\bm k$ following absorption and emission of a phonon with wave vector $\bm q$. The rate of variation of the electron distribution function mediated by interaction with phonons can is represented in Eq.~\eqref{4types} in full form in the standard way through the phonon collision integral \cite{Anselm}.
\begin{widetext}
\begin{multline}\label{4types}
\left(\frac{\partial}{\partial t} f(\bm k) \right)_{ph} = - \int{\frac{d \bm q}{4 \pi^2}}
\left[
W_{\bm k \rightarrow \bm k + \bm q}^{abs}f^{(0)}(\varepsilon_{\bm k}) \left(1 - f^{(0)}(\varepsilon_{\bm k + \bm q}) \right) + W_{\bm k \rightarrow \bm k - \bm q}^{em}f^{(0)}(\varepsilon_{\bm k}) \left(1 - f^{(0)}(\varepsilon_{\bm k - \bm q}) \right) \right. \\
-\left. W_{\bm k - \bm q \rightarrow \bm k}^{abs}f^{(0)}(\varepsilon_{\bm k - \bm q}) \left(1 - f^{(0)}(\varepsilon_{\bm k }) \right) - W_{\bm k + \bm q \rightarrow \bm k}^{em}f^{(0)}(\varepsilon_{\bm k + \bm q}) \left(1 - f^{(0)}(\varepsilon_{\bm k})
\right) \right]\,,
\end{multline}
\end{widetext}

In Eq. \eqref{4types} $f^{(0)}(\varepsilon_{\bm k})$ is the equilibrium distribution function of electrons with a chemical potential depending on $\varepsilon_{F}$ and temperature. At low temperatures, the chemical potential can be assumed equal to  $\varepsilon_{F}$, and the correction accounting for temperature variation may be neglected. Integration is run over the phonon wave vector $\bm q$ over the whole Brillouin zone. Scattering from acoustic phonons does not initiate interband and intervalley transitions and variation of the electron spin.

Obviously enough, an equilibrium distribution function describes randomized phonons, and drag is mediated only by the nonequilibrium term $N_{ph}^{(1)}$ which should be substituted from \eqref{correction_phon} into \eqref{Fermi_golden}, and further, into \eqref{4types}. $v_F$ being approximately 50 times larger than $v_s$, electrons scatter from phonons in an almost elastic way. The energy conservation law is fundamentally important in including into \eqref{4types} the distribution function of electrons before and after their scattering from a phonon \cite{Anselm}. One can set
\begin{equation} \label{inelasticy}
f^{(0)}(\bm k + \bm q) = f^{(0)}(\bm k ) \pm \frac{\partial f^{0}(\varepsilon)}{\partial \varepsilon}\hbar \omega_{ph}(\bm q)\,.
\end{equation}
The "+" sign identifies here absorption of a quantum of sound, and the "-" sign, emission. Conversely, integrating in angle, one can set
\begin{equation} \label{elasticy}
\delta \left( k^\prime - k \pm \frac{\omega_{ph} (\bm q)}{v_F} \right)
\approx
\frac{1}{q}\delta \left(  \cos{\theta_{\bm k, \bm q} + \frac{q}{2k}}  \right)\,,
\end{equation}
which corresponds to neglecting phonon energy in a considering of scattering. As a result, interaction with a phonon flux transforms the collision integral to
\begin{equation} \label{St_final}
\left(\frac{\partial}{\partial t} f(\bm k) \right)_{ph} =
v_F \frac{\bm k}{k} k_B\nabla T G_{ph}(k) \frac{\partial f^{(0)}}{\partial \varepsilon}\,,
\end{equation}
where the dimensionless parameter is introduced
\begin{equation} \label{G}
G_{ph}(k)=\frac{S}{2M N_u} \frac{D^2}{\hbar^3 v_F^4} \frac{L}{v_s} \varepsilon^2(k).
\end{equation}

We readily see that the phonon collision integral includes the quantity $ \frac{\partial f^{(0)}}{\partial \varepsilon} $ and a scalar product of the electron wave vector $\bm k$ by the vector describing external factor $\nabla T$. Hence, the phonon collision integral can be interpreted as a field term of the Boltzmann kinetic equation. This makes possible calculation of the nonequilibrium term $f^{(1)}(\bm k)$ for the electron distribution function $f^{(0)}(\varepsilon_{\bm k})$ in the relaxation time approximation by straightforward multiplication of the right-hand part of relation relation ~\eqref{St_final} by the electron transport relaxation time $\tau(\varepsilon_{\bm k})$.

One can now calculate the electric current, as well as the current $\bm j$ generated by the external electric field. In derivation of the current $\bm j =\frac{e}{\pi^2} \int{d\bm k\, \bm v_{\bm k} f^{(1)}(\bm k)}$, the first approximation (delta-function approximation) in relation \eqref{inelasticy} for $\frac{\partial f^{(0)}}{\partial \varepsilon}$ will be already sufficient. The final expression for the contribution provided by the phonon drag to thermopower looks as

\begin{equation} \label{S_final}
S_{ph}=\frac{ \pi^4}{2e} \frac{k_B\hbar }{M a_0^2}\frac{L}{v_s}
\left( \frac{D}{\hbar v_F \pi / a_0} \right)^2
\left( \frac{\varepsilon_F}{\hbar v_F \pi / a_0} \right)^2.
\end{equation}

The transport electron relaxation time cancels when current is divided by the electrical conductivity of the sample.

The most remarkable feature in the relation obtained is that it does not contain an explicit dependence of thermopower on temperature, which argues with \cite{PSSB:PSSB201248302}. The obtained relation correlates with the observation that only a directed phonon flux (phonon wind) contributes to thermopower; now, in accordance with relation ~\eqref{correction_phon}, this flux is proportional to temperature gradient and does not depend directly on the temperature itself.

The non-dimensional factor $\frac{\hbar \tau_{ph}}{M a_0^2}$ in. relation ~\eqref{S_final} is close in magnitude to 8, and and the ratio $\frac{D}{\hbar v_F \pi / a_0}$ approaches 2.

The obtained phonon drag thermopower should be compared with well known formula (see for instance relation (4) from Ref. \cite{PhysRevB.80.235415}) of diffusion thermopower in graphene:
\begin{equation} \label{S_diff}
S_{d} = \frac{\pi^2}{3e} \frac{k_B^2T}{\varepsilon_F}.
\end{equation}

In the case of a degenerate electron gas, the diffusion thermoelectric current turns out to be zero in the first approximation for $\frac{\partial f^{(0)}}{\partial \varepsilon}$. Taking the second term in the Sommerfeld expansion results in the proportionality to $T/\varepsilon_F$, which figures in \eqref{S_diff}. This expression for the diffusion factor in the thermopower in graphene correlates with the Mott formula and coincides with that developed for metals \cite{Dau10en} and graphite \cite{PhysRevB.34.4298}. To bring the result in compliance with the mechanism of carrier scattering in a particular sample, one has naturally to introduce a factor of order unity.

\begin{figure}
\includegraphics[width=1.0\linewidth]{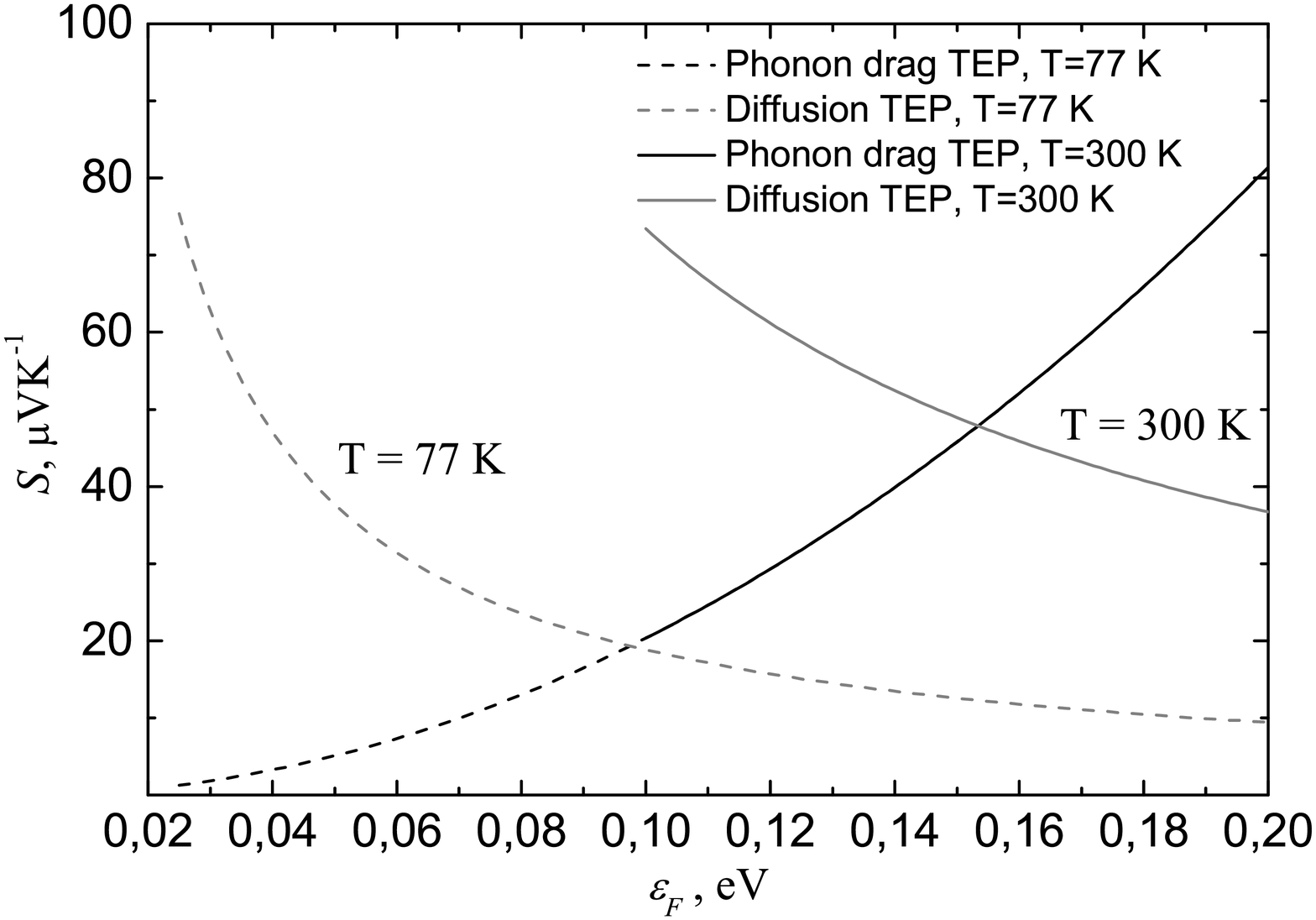}
\caption{Two contributions to the Seebeck coefficient for a sample 2\,$\mu$m in size and regime of phonon scattering at the boundary plotted as a function of Fermi energy. Solid curves relate to room temperature, more specifically, the solid black line visualizes the phonon drag contribution $S_ph$, and the solid grey line, the diffusion contribution $S_d$. These graphs start at $0.1\mathrm{eV}$, where the degenerate electron gas approximation holds. The dotted black line depicts the phonon drag contribution, and the dotted grey line, the diffusion contribution at $T = 77$\,K.}
\label{fig2}
\end{figure}

Figure \ref{fig2} plots the dependencies of the contribution to thermopower due to electron drag by phonons and of the contribution to thermopower due to diffusion, as functions of the Fermi level, which were calculated by relations~\eqref{S_final} and \eqref{S_diff}, respectively.

In most of the available relevant publications \cite{doi:10.1021/nl102923q,PhysRevB.80.081413,PhysRevLett.102.166808,PhysRevLett.102.096807,PhysRevB.83.113403} thermoelectric properties of graphene find straightforward interpretation in terms of the Mott formula. No phonon drag was observed in these studies. Small deviations from the Mott formula were recorded, however, at low temperatures. These deviations should most probably be assigned only to a decrease of measurement accuracy. Only the data reported in Ref. \cite{doi:10.1021/nl102923q} may be interpreted as reflecting saturation of thermopower with decreasing temperature. Finally, several graphs in Ref. \cite{PhysRevLett.102.096807} suggest growth of thermopower with increasing Fermi level.

The first reason for the phonon drag effect becoming suppressed could be the reverse effect exerted by electrons on the phonon system, accompanied by the corresponding decrease of the phonon relaxation time \cite{Anselm} The second possible factor could be the effect of the substrate, which may likewise reduce the phonon mean free paths. Intrinsic phonons of graphene may be expected to become scattered from phonons of the substrate or from substrate irregularities. For instance, the limitation of the electron mobility caused by the phonons from the substrate was indicated in \cite{SiO2Limit}. Molecular dynamics modeling \cite{MDphonons} yields a value of about $5 \mathrm{ps}$ for the relaxation time of long-wavelength acoustic phonons in  $4 \times 4 \mathrm{nm^2}$ graphene sheets on a SiO$_2$ substrate. For a free-standing sample, the relaxation time of such phonons turned out to be about 10 times larger, approaching in magnitude the value of 100 ps used in this study.  Nevertheless, it appears pertinent to note that interaction of mechanically exfoliated graphene with substrate is assigned to weak van-der-Waals forces \cite{GrapheneSiO2,GrapheneIr111}. Substrate, in particular, does not affect the shape of Raman spectra \cite{doi:10.1021/jp8008404}, which are very sensitive probes of the vibrational characteristics of a crystal.

A calculation was made of the phonon drag contribution to thermopower in graphene in the Bloch-Gr\"{u}neisen (BG) temperature regime\cite{PhysRevB.79.075417}. The BG regime specifies the case where temperature is not high enough to drive excitation of phonons with a wave vector on the order of the characteristic electron wave vector. The Seebeck coefficient in this case is proportional to $T^3$, the situation coinciding with the case of graphite \cite{PhysRevB.34.4298}. To realize the BG regime in graphene, however, the temperatures should be at their lowest limit. As already mentioned, in the case of a degenerate electron gas phonons interacting with electrons of energy on the order of $\varepsilon_F \sim 0.1 \mathrm{eV}$ have a a temperature of about 20\,K. The very noticeable predictions of thermopower enhancement are made for graphene with band gap opened for both diffusion \cite{PhysRevB.86.075411} and phonon drag contributions \cite{PhysRevB.82.155410}.

As it was mentioned above, the phonons that generate the drag of electrons are very small part of the excited phonons that define thermal conductivity. In theory it allows to manipulate the thermal conductivity and the phonon drag thermopower separately. We suggest that the phonon drag thermopower is defined by the scattering of phonons on the edges of the sample. Thermal conductivity can also rise nearly up to the room temperature with increasing the sample size. But due to relation $ZT=\sigma T S^2/\kappa$ for thermoelectric figure of merit it will grow with increasing of sample size linearly or even faster. The problem of the thermal conductivity reducing is complicated because making samples with point defects, placing samples onto the substrate or increasing ambient temperature will reduce both electrical and thermal conductivity of the sample and these effects should be considered simultaneously. Thereby tuning the size of the sample seems to be the straightest way of improving graphene thermoelectrical properties.

The quality of graphene samples prepared is improving continuously \cite{RevModPhys.83.407,Dideykin2011105,doi:10.1021/nl072566s}. As demonstrated in this study, in a high quality sample with a large phonon mean free path, which probably correlates with high thermal conductivity, at temperatures of the environment from liquid nitrogen to room temperature, and at a high Fermi level one may expect the contribution to thermopower due to phonon drag to dominate over the diffusion-associated one, and a sharp increase of the Seebeck coefficient and thermoelectric figure of merit.

\acknowledgments
We are deeply grateful to M.M. Glazov for support and fruitful discussions. Our thanks are due to A.Ya. Vul' and A.T. Dideykin for interest to work. This study was supported by a grant for young scientists of the Dynasty foundation. Partial support was provided by the Russian Ministry of Education and Science (projects 8683).

\begin{thebibliography}{46}%
\makeatletter
\providecommand \@ifxundefined [1]{%
 \@ifx{#1\undefined}
}%
\providecommand \@ifnum [1]{%
 \ifnum #1\expandafter \@firstoftwo
 \else \expandafter \@secondoftwo
 \fi
}%
\providecommand \@ifx [1]{%
 \ifx #1\expandafter \@firstoftwo
 \else \expandafter \@secondoftwo
 \fi
}%
\providecommand \natexlab [1]{#1}%
\providecommand \enquote  [1]{``#1''}%
\providecommand \bibnamefont  [1]{#1}%
\providecommand \bibfnamefont [1]{#1}%
\providecommand \citenamefont [1]{#1}%
\providecommand \href@noop [0]{\@secondoftwo}%
\providecommand \href [0]{\begingroup \@sanitize@url \@href}%
\providecommand \@href[1]{\@@startlink{#1}\@@href}%
\providecommand \@@href[1]{\endgroup#1\@@endlink}%
\providecommand \@sanitize@url [0]{\catcode `\\12\catcode `\$12\catcode
  `\&12\catcode `\#12\catcode `\^12\catcode `\_12\catcode `\%12\relax}%
\providecommand \@@startlink[1]{}%
\providecommand \@@endlink[0]{}%
\providecommand \url  [0]{\begingroup\@sanitize@url \@url }%
\providecommand \@url [1]{\endgroup\@href {#1}{\urlprefix }}%
\providecommand \urlprefix  [0]{URL }%
\providecommand \Eprint [0]{\href }%
\providecommand \doibase [0]{http://dx.doi.org/}%
\providecommand \selectlanguage [0]{\@gobble}%
\providecommand \bibinfo  [0]{\@secondoftwo}%
\providecommand \bibfield  [0]{\@secondoftwo}%
\providecommand \translation [1]{[#1]}%
\providecommand \BibitemOpen [0]{}%
\providecommand \bibitemStop [0]{}%
\providecommand \bibitemNoStop [0]{.\EOS\space}%
\providecommand \EOS [0]{\spacefactor3000\relax}%
\providecommand \BibitemShut  [1]{\csname bibitem#1\endcsname}%
\let\auto@bib@innerbib\@empty
\bibitem [{\citenamefont {Castro~Neto}\ \emph {et~al.}(2009)\citenamefont
  {Castro~Neto}, \citenamefont {Guinea}, \citenamefont {Peres}, \citenamefont
  {Novoselov},\ and\ \citenamefont {Geim}}]{RevModPhys.81.109}%
  \BibitemOpen
  \bibfield  {author} {\bibinfo {author} {\bibfnamefont {A.~H.}\ \bibnamefont
  {Castro~Neto}}, \bibinfo {author} {\bibfnamefont {F.}~\bibnamefont {Guinea}},
  \bibinfo {author} {\bibfnamefont {N.~M.~R.}\ \bibnamefont {Peres}}, \bibinfo
  {author} {\bibfnamefont {K.~S.}\ \bibnamefont {Novoselov}}, \ and\ \bibinfo
  {author} {\bibfnamefont {A.~K.}\ \bibnamefont {Geim}},\ }\href {\doibase
  10.1103/RevModPhys.81.109} {\bibfield  {journal} {\bibinfo  {journal} {Rev.
  Mod. Phys.}\ }\textbf {\bibinfo {volume} {81}},\ \bibinfo {pages} {109}
  (\bibinfo {year} {2009})}\BibitemShut {NoStop}%
\bibitem [{\citenamefont {Das~Sarma}\ \emph {et~al.}(2011)\citenamefont
  {Das~Sarma}, \citenamefont {Adam}, \citenamefont {Hwang},\ and\ \citenamefont
  {Rossi}}]{RevModPhys.83.407}%
  \BibitemOpen
  \bibfield  {author} {\bibinfo {author} {\bibfnamefont {S.}~\bibnamefont
  {Das~Sarma}}, \bibinfo {author} {\bibfnamefont {S.}~\bibnamefont {Adam}},
  \bibinfo {author} {\bibfnamefont {E.~H.}\ \bibnamefont {Hwang}}, \ and\
  \bibinfo {author} {\bibfnamefont {E.}~\bibnamefont {Rossi}},\ }\href
  {\doibase 10.1103/RevModPhys.83.407} {\bibfield  {journal} {\bibinfo
  {journal} {Rev. Mod. Phys.}\ }\textbf {\bibinfo {volume} {83}},\ \bibinfo
  {pages} {407} (\bibinfo {year} {2011})}\BibitemShut {NoStop}%
\bibitem [{\citenamefont {Balandin}(2011)}]{BalReview}%
  \BibitemOpen
  \bibfield  {author} {\bibinfo {author} {\bibfnamefont {A.~A.}\ \bibnamefont
  {Balandin}},\ }\href {\doibase http://dx.doi.org/10.1038/nmat3064} {\bibfield
   {journal} {\bibinfo  {journal} {Journal of Physics: Condensed Matter}\
  }\textbf {\bibinfo {volume} {10}},\ \bibinfo {pages} {560} (\bibinfo {year}
  {2011})}\BibitemShut {NoStop}%
\bibitem [{\citenamefont {Nika}\ and\ \citenamefont
  {Balandin}(2012)}]{NikaBalReview}%
  \BibitemOpen
  \bibfield  {author} {\bibinfo {author} {\bibfnamefont {D.~L.}\ \bibnamefont
  {Nika}}\ and\ \bibinfo {author} {\bibfnamefont {A.~A.}\ \bibnamefont
  {Balandin}},\ }\href {http://stacks.iop.org/0953-8984/24/i=23/a=233203}
  {\bibfield  {journal} {\bibinfo  {journal} {Journal of Physics: Condensed
  Matter}\ }\textbf {\bibinfo {volume} {24}},\ \bibinfo {pages} {233203}
  (\bibinfo {year} {2012})}\BibitemShut {NoStop}%
\bibitem [{\citenamefont {Mariani}\ and\ \citenamefont {von
  Oppen}(2010)}]{PhysRevB.82.195403}%
  \BibitemOpen
  \bibfield  {author} {\bibinfo {author} {\bibfnamefont {E.}~\bibnamefont
  {Mariani}}\ and\ \bibinfo {author} {\bibfnamefont {F.}~\bibnamefont {von
  Oppen}},\ }\href {\doibase 10.1103/PhysRevB.82.195403} {\bibfield  {journal}
  {\bibinfo  {journal} {Phys. Rev. B}\ }\textbf {\bibinfo {volume} {82}},\
  \bibinfo {pages} {195403} (\bibinfo {year} {2010})}\BibitemShut {NoStop}%
\bibitem [{\citenamefont {Efetov}\ and\ \citenamefont
  {Kim}(2010)}]{PhysRevLett.105.256805}%
  \BibitemOpen
  \bibfield  {author} {\bibinfo {author} {\bibfnamefont {D.~K.}\ \bibnamefont
  {Efetov}}\ and\ \bibinfo {author} {\bibfnamefont {P.}~\bibnamefont {Kim}},\
  }\href {\doibase 10.1103/PhysRevLett.105.256805} {\bibfield  {journal}
  {\bibinfo  {journal} {Phys. Rev. Lett.}\ }\textbf {\bibinfo {volume} {105}},\
  \bibinfo {pages} {256805} (\bibinfo {year} {2010})}\BibitemShut {NoStop}%
\bibitem [{\citenamefont {Alofi}\ and\ \citenamefont
  {Srivastava}(2012)}]{6235226}%
  \BibitemOpen
  \bibfield  {author} {\bibinfo {author} {\bibfnamefont {A.}~\bibnamefont
  {Alofi}}\ and\ \bibinfo {author} {\bibfnamefont {G.~P.}\ \bibnamefont
  {Srivastava}},\ }\href {\doibase 10.1063/1.4733690} {\bibfield  {journal}
  {\bibinfo  {journal} {Journal of Applied Physics}\ }\textbf {\bibinfo
  {volume} {112}},\ \bibinfo {pages} {013517 } (\bibinfo {year}
  {2012})}\BibitemShut {NoStop}%
\bibitem [{\citenamefont {Nika}\ \emph {et~al.}(2009)\citenamefont {Nika},
  \citenamefont {Pokatilov}, \citenamefont {Askerov},\ and\ \citenamefont
  {Balandin}}]{PhysRevB.79.155413}%
  \BibitemOpen
  \bibfield  {author} {\bibinfo {author} {\bibfnamefont {D.~L.}\ \bibnamefont
  {Nika}}, \bibinfo {author} {\bibfnamefont {E.~P.}\ \bibnamefont {Pokatilov}},
  \bibinfo {author} {\bibfnamefont {A.~S.}\ \bibnamefont {Askerov}}, \ and\
  \bibinfo {author} {\bibfnamefont {A.~A.}\ \bibnamefont {Balandin}},\ }\href
  {\doibase 10.1103/PhysRevB.79.155413} {\bibfield  {journal} {\bibinfo
  {journal} {Phys. Rev. B}\ }\textbf {\bibinfo {volume} {79}},\ \bibinfo
  {pages} {155413} (\bibinfo {year} {2009})}\BibitemShut {NoStop}%
\bibitem [{\citenamefont {Seol}\ \emph {et~al.}(2010)\citenamefont {Seol},
  \citenamefont {Jo}, \citenamefont {Moore}, \citenamefont {Lindsay},
  \citenamefont {Aitken}, \citenamefont {Pettes}, \citenamefont {Li},
  \citenamefont {Yao}, \citenamefont {Huang}, \citenamefont {Broido},
  \citenamefont {Mingo}, \citenamefont {Ruoff},\ and\ \citenamefont
  {Shi}}]{Seol09042010}%
  \BibitemOpen
  \bibfield  {author} {\bibinfo {author} {\bibfnamefont {J.~H.}\ \bibnamefont
  {Seol}}, \bibinfo {author} {\bibfnamefont {I.}~\bibnamefont {Jo}}, \bibinfo
  {author} {\bibfnamefont {A.~L.}\ \bibnamefont {Moore}}, \bibinfo {author}
  {\bibfnamefont {L.}~\bibnamefont {Lindsay}}, \bibinfo {author} {\bibfnamefont
  {Z.~H.}\ \bibnamefont {Aitken}}, \bibinfo {author} {\bibfnamefont {M.~T.}\
  \bibnamefont {Pettes}}, \bibinfo {author} {\bibfnamefont {X.}~\bibnamefont
  {Li}}, \bibinfo {author} {\bibfnamefont {Z.}~\bibnamefont {Yao}}, \bibinfo
  {author} {\bibfnamefont {R.}~\bibnamefont {Huang}}, \bibinfo {author}
  {\bibfnamefont {D.}~\bibnamefont {Broido}}, \bibinfo {author} {\bibfnamefont
  {N.}~\bibnamefont {Mingo}}, \bibinfo {author} {\bibfnamefont {R.~S.}\
  \bibnamefont {Ruoff}}, \ and\ \bibinfo {author} {\bibfnamefont
  {L.}~\bibnamefont {Shi}},\ }\href {\doibase 10.1126/science.1184014}
  {\bibfield  {journal} {\bibinfo  {journal} {Science}\ }\textbf {\bibinfo
  {volume} {328}},\ \bibinfo {pages} {213} (\bibinfo {year} {2010})},\ \Eprint
  {http://arxiv.org/abs/http://www.sciencemag.org/content/328/5975/213.full.pdf}
  {http://www.sciencemag.org/content/328/5975/213.full.pdf} \BibitemShut
  {NoStop}%
\bibitem [{\citenamefont {Wang}\ \emph {et~al.}(2011)\citenamefont {Wang},
  \citenamefont {Xie}, \citenamefont {Bui}, \citenamefont {Liu}, \citenamefont
  {Ni}, \citenamefont {Li},\ and\ \citenamefont
  {Thong}}]{doi:10.1021/nl102923q}%
  \BibitemOpen
  \bibfield  {author} {\bibinfo {author} {\bibfnamefont {Z.}~\bibnamefont
  {Wang}}, \bibinfo {author} {\bibfnamefont {R.}~\bibnamefont {Xie}}, \bibinfo
  {author} {\bibfnamefont {C.~T.}\ \bibnamefont {Bui}}, \bibinfo {author}
  {\bibfnamefont {D.}~\bibnamefont {Liu}}, \bibinfo {author} {\bibfnamefont
  {X.}~\bibnamefont {Ni}}, \bibinfo {author} {\bibfnamefont {B.}~\bibnamefont
  {Li}}, \ and\ \bibinfo {author} {\bibfnamefont {J.~T.~L.}\ \bibnamefont
  {Thong}},\ }\href {\doibase 10.1021/nl102923q} {\bibfield  {journal}
  {\bibinfo  {journal} {Nano Letters}\ }\textbf {\bibinfo {volume} {11}},\
  \bibinfo {pages} {113} (\bibinfo {year} {2011})},\ \Eprint
  {http://arxiv.org/abs/http://pubs.acs.org/doi/pdf/10.1021/nl102923q}
  {http://pubs.acs.org/doi/pdf/10.1021/nl102923q} \BibitemShut {NoStop}%
\bibitem [{\citenamefont {Luk'yanchuk}\ \emph {et~al.}(2011)\citenamefont
  {Luk'yanchuk}, \citenamefont {Varlamov},\ and\ \citenamefont
  {Kavokin}}]{PhysRevLett.107.016601}%
  \BibitemOpen
  \bibfield  {author} {\bibinfo {author} {\bibfnamefont {I.~A.}\ \bibnamefont
  {Luk'yanchuk}}, \bibinfo {author} {\bibfnamefont {A.~A.}\ \bibnamefont
  {Varlamov}}, \ and\ \bibinfo {author} {\bibfnamefont {A.~V.}\ \bibnamefont
  {Kavokin}},\ }\href {\doibase 10.1103/PhysRevLett.107.016601} {\bibfield
  {journal} {\bibinfo  {journal} {Phys. Rev. Lett.}\ }\textbf {\bibinfo
  {volume} {107}},\ \bibinfo {pages} {016601} (\bibinfo {year}
  {2011})}\BibitemShut {NoStop}%
\bibitem [{\citenamefont {Nalitov}\ \emph {et~al.}(2012)\citenamefont
  {Nalitov}, \citenamefont {Golub},\ and\ \citenamefont
  {Ivchenko}}]{PhysRevB.86.115301}%
  \BibitemOpen
  \bibfield  {author} {\bibinfo {author} {\bibfnamefont {A.~V.}\ \bibnamefont
  {Nalitov}}, \bibinfo {author} {\bibfnamefont {L.~E.}\ \bibnamefont {Golub}},
  \ and\ \bibinfo {author} {\bibfnamefont {E.~L.}\ \bibnamefont {Ivchenko}},\
  }\href {\doibase 10.1103/PhysRevB.86.115301} {\bibfield  {journal} {\bibinfo
  {journal} {Phys. Rev. B}\ }\textbf {\bibinfo {volume} {86}},\ \bibinfo
  {pages} {115301} (\bibinfo {year} {2012})}\BibitemShut {NoStop}%
\bibitem [{\citenamefont {Jiang}\ \emph {et~al.}(2011)\citenamefont {Jiang},
  \citenamefont {Shalygin}, \citenamefont {Panevin}, \citenamefont {Danilov},
  \citenamefont {Glazov}, \citenamefont {Yakimova}, \citenamefont {Lara-Avila},
  \citenamefont {Kubatkin},\ and\ \citenamefont
  {Ganichev}}]{PhysRevB.84.125429}%
  \BibitemOpen
  \bibfield  {author} {\bibinfo {author} {\bibfnamefont {C.}~\bibnamefont
  {Jiang}}, \bibinfo {author} {\bibfnamefont {V.~A.}\ \bibnamefont {Shalygin}},
  \bibinfo {author} {\bibfnamefont {V.~Y.}\ \bibnamefont {Panevin}}, \bibinfo
  {author} {\bibfnamefont {S.~N.}\ \bibnamefont {Danilov}}, \bibinfo {author}
  {\bibfnamefont {M.~M.}\ \bibnamefont {Glazov}}, \bibinfo {author}
  {\bibfnamefont {R.}~\bibnamefont {Yakimova}}, \bibinfo {author}
  {\bibfnamefont {S.}~\bibnamefont {Lara-Avila}}, \bibinfo {author}
  {\bibfnamefont {S.}~\bibnamefont {Kubatkin}}, \ and\ \bibinfo {author}
  {\bibfnamefont {S.~D.}\ \bibnamefont {Ganichev}},\ }\href {\doibase
  10.1103/PhysRevB.84.125429} {\bibfield  {journal} {\bibinfo  {journal} {Phys.
  Rev. B}\ }\textbf {\bibinfo {volume} {84}},\ \bibinfo {pages} {125429}
  (\bibinfo {year} {2011})}\BibitemShut {NoStop}%
\bibitem [{\citenamefont {Checkelsky}\ and\ \citenamefont
  {Ong}(2009)}]{PhysRevB.80.081413}%
  \BibitemOpen
  \bibfield  {author} {\bibinfo {author} {\bibfnamefont {J.~G.}\ \bibnamefont
  {Checkelsky}}\ and\ \bibinfo {author} {\bibfnamefont {N.~P.}\ \bibnamefont
  {Ong}},\ }\href {\doibase 10.1103/PhysRevB.80.081413} {\bibfield  {journal}
  {\bibinfo  {journal} {Phys. Rev. B}\ }\textbf {\bibinfo {volume} {80}},\
  \bibinfo {pages} {081413} (\bibinfo {year} {2009})}\BibitemShut {NoStop}%
\bibitem [{\citenamefont {Wei}\ \emph {et~al.}(2009)\citenamefont {Wei},
  \citenamefont {Bao}, \citenamefont {Pu}, \citenamefont {Lau},\ and\
  \citenamefont {Shi}}]{PhysRevLett.102.166808}%
  \BibitemOpen
  \bibfield  {author} {\bibinfo {author} {\bibfnamefont {P.}~\bibnamefont
  {Wei}}, \bibinfo {author} {\bibfnamefont {W.}~\bibnamefont {Bao}}, \bibinfo
  {author} {\bibfnamefont {Y.}~\bibnamefont {Pu}}, \bibinfo {author}
  {\bibfnamefont {C.~N.}\ \bibnamefont {Lau}}, \ and\ \bibinfo {author}
  {\bibfnamefont {J.}~\bibnamefont {Shi}},\ }\href {\doibase
  10.1103/PhysRevLett.102.166808} {\bibfield  {journal} {\bibinfo  {journal}
  {Phys. Rev. Lett.}\ }\textbf {\bibinfo {volume} {102}},\ \bibinfo {pages}
  {166808} (\bibinfo {year} {2009})}\BibitemShut {NoStop}%
\bibitem [{\citenamefont {Zuev}\ \emph {et~al.}(2009)\citenamefont {Zuev},
  \citenamefont {Chang},\ and\ \citenamefont {Kim}}]{PhysRevLett.102.096807}%
  \BibitemOpen
  \bibfield  {author} {\bibinfo {author} {\bibfnamefont {Y.~M.}\ \bibnamefont
  {Zuev}}, \bibinfo {author} {\bibfnamefont {W.}~\bibnamefont {Chang}}, \ and\
  \bibinfo {author} {\bibfnamefont {P.}~\bibnamefont {Kim}},\ }\href {\doibase
  10.1103/PhysRevLett.102.096807} {\bibfield  {journal} {\bibinfo  {journal}
  {Phys. Rev. Lett.}\ }\textbf {\bibinfo {volume} {102}},\ \bibinfo {pages}
  {096807} (\bibinfo {year} {2009})}\BibitemShut {NoStop}%
\bibitem [{\citenamefont {Wang}\ and\ \citenamefont
  {Shi}(2011)}]{PhysRevB.83.113403}%
  \BibitemOpen
  \bibfield  {author} {\bibinfo {author} {\bibfnamefont {D.}~\bibnamefont
  {Wang}}\ and\ \bibinfo {author} {\bibfnamefont {J.}~\bibnamefont {Shi}},\
  }\href {\doibase 10.1103/PhysRevB.83.113403} {\bibfield  {journal} {\bibinfo
  {journal} {Phys. Rev. B}\ }\textbf {\bibinfo {volume} {83}},\ \bibinfo
  {pages} {113403} (\bibinfo {year} {2011})}\BibitemShut {NoStop}%
\bibitem [{\citenamefont {Gabor}\ \emph {et~al.}(2011)\citenamefont {Gabor},
  \citenamefont {Song}, \citenamefont {Ma}, \citenamefont {Nair}, \citenamefont
  {Taychatanapat}, \citenamefont {Watanabe}, \citenamefont {Taniguchi},
  \citenamefont {Levitov},\ and\ \citenamefont
  {Jarillo-Herrero}}]{Gabor04112011}%
  \BibitemOpen
  \bibfield  {author} {\bibinfo {author} {\bibfnamefont {N.~M.}\ \bibnamefont
  {Gabor}}, \bibinfo {author} {\bibfnamefont {J.~C.~W.}\ \bibnamefont {Song}},
  \bibinfo {author} {\bibfnamefont {Q.}~\bibnamefont {Ma}}, \bibinfo {author}
  {\bibfnamefont {N.~L.}\ \bibnamefont {Nair}}, \bibinfo {author}
  {\bibfnamefont {T.}~\bibnamefont {Taychatanapat}}, \bibinfo {author}
  {\bibfnamefont {K.}~\bibnamefont {Watanabe}}, \bibinfo {author}
  {\bibfnamefont {T.}~\bibnamefont {Taniguchi}}, \bibinfo {author}
  {\bibfnamefont {L.~S.}\ \bibnamefont {Levitov}}, \ and\ \bibinfo {author}
  {\bibfnamefont {P.}~\bibnamefont {Jarillo-Herrero}},\ }\href {\doibase
  10.1126/science.1211384} {\bibfield  {journal} {\bibinfo  {journal}
  {Science}\ }\textbf {\bibinfo {volume} {334}},\ \bibinfo {pages} {648}
  (\bibinfo {year} {2011})},\ \Eprint
  {http://arxiv.org/abs/http://www.sciencemag.org/content/334/6056/648.full.pdf}
  {http://www.sciencemag.org/content/334/6056/648.full.pdf} \BibitemShut
  {NoStop}%
\bibitem [{\citenamefont {Kubakaddi}(2009)}]{PhysRevB.79.075417}%
  \BibitemOpen
  \bibfield  {author} {\bibinfo {author} {\bibfnamefont {S.~S.}\ \bibnamefont
  {Kubakaddi}},\ }\href {\doibase 10.1103/PhysRevB.79.075417} {\bibfield
  {journal} {\bibinfo  {journal} {Phys. Rev. B}\ }\textbf {\bibinfo {volume}
  {79}},\ \bibinfo {pages} {075417} (\bibinfo {year} {2009})}\BibitemShut
  {NoStop}%
\bibitem [{\citenamefont {Kubakaddi}\ and\ \citenamefont
  {Bhargavi}(2010)}]{PhysRevB.82.155410}%
  \BibitemOpen
  \bibfield  {author} {\bibinfo {author} {\bibfnamefont {S.~S.}\ \bibnamefont
  {Kubakaddi}}\ and\ \bibinfo {author} {\bibfnamefont {K.~S.}\ \bibnamefont
  {Bhargavi}},\ }\href {\doibase 10.1103/PhysRevB.82.155410} {\bibfield
  {journal} {\bibinfo  {journal} {Phys. Rev. B}\ }\textbf {\bibinfo {volume}
  {82}},\ \bibinfo {pages} {155410} (\bibinfo {year} {2010})}\BibitemShut
  {NoStop}%
\bibitem [{\citenamefont {Hwang}\ \emph {et~al.}(2009)\citenamefont {Hwang},
  \citenamefont {Rossi},\ and\ \citenamefont {Das~Sarma}}]{PhysRevB.80.235415}%
  \BibitemOpen
  \bibfield  {author} {\bibinfo {author} {\bibfnamefont {E.~H.}\ \bibnamefont
  {Hwang}}, \bibinfo {author} {\bibfnamefont {E.}~\bibnamefont {Rossi}}, \ and\
  \bibinfo {author} {\bibfnamefont {S.}~\bibnamefont {Das~Sarma}},\ }\href
  {\doibase 10.1103/PhysRevB.80.235415} {\bibfield  {journal} {\bibinfo
  {journal} {Phys. Rev. B}\ }\textbf {\bibinfo {volume} {80}},\ \bibinfo
  {pages} {235415} (\bibinfo {year} {2009})}\BibitemShut {NoStop}%
\bibitem [{\citenamefont {Vaidya}\ \emph {et~al.}(2012)\citenamefont {Vaidya},
  \citenamefont {Sankeshwar},\ and\ \citenamefont {Mulimani}}]{vaidya:093711}%
  \BibitemOpen
  \bibfield  {author} {\bibinfo {author} {\bibfnamefont {R.~G.}\ \bibnamefont
  {Vaidya}}, \bibinfo {author} {\bibfnamefont {N.~S.}\ \bibnamefont
  {Sankeshwar}}, \ and\ \bibinfo {author} {\bibfnamefont {B.~G.}\ \bibnamefont
  {Mulimani}},\ }\href {\doibase 10.1063/1.4764335} {\bibfield  {journal}
  {\bibinfo  {journal} {Journal of Applied Physics}\ }\textbf {\bibinfo
  {volume} {112}},\ \bibinfo {eid} {093711} (\bibinfo {year}
  {2012})}\BibitemShut {NoStop}%
\bibitem [{\citenamefont {Sankeshwar}\ \emph {et~al.}(2013)\citenamefont
  {Sankeshwar}, \citenamefont {Vaidya},\ and\ \citenamefont
  {Mulimani}}]{PSSB:PSSB201248302}%
  \BibitemOpen
  \bibfield  {author} {\bibinfo {author} {\bibfnamefont {N.~S.}\ \bibnamefont
  {Sankeshwar}}, \bibinfo {author} {\bibfnamefont {R.~G.}\ \bibnamefont
  {Vaidya}}, \ and\ \bibinfo {author} {\bibfnamefont {B.~G.}\ \bibnamefont
  {Mulimani}},\ }\href {\doibase 10.1002/pssb.201248302} {\bibfield  {journal}
  {\bibinfo  {journal} {physica status solidi (b)}\ }\textbf {\bibinfo {volume}
  {250}},\ \bibinfo {pages} {1356} (\bibinfo {year} {2013})}\BibitemShut
  {NoStop}%
\bibitem [{\citenamefont {Patel}\ and\ \citenamefont
  {Mukerjee}(2012)}]{PhysRevB.86.075411}%
  \BibitemOpen
  \bibfield  {author} {\bibinfo {author} {\bibfnamefont {A.~A.}\ \bibnamefont
  {Patel}}\ and\ \bibinfo {author} {\bibfnamefont {S.}~\bibnamefont
  {Mukerjee}},\ }\href {\doibase 10.1103/PhysRevB.86.075411} {\bibfield
  {journal} {\bibinfo  {journal} {Phys. Rev. B}\ }\textbf {\bibinfo {volume}
  {86}},\ \bibinfo {pages} {075411} (\bibinfo {year} {2012})}\BibitemShut
  {NoStop}%
\bibitem [{\citenamefont {Tse}\ and\ \citenamefont
  {Das~Sarma}(2009)}]{PhysRevB.79.235406}%
  \BibitemOpen
  \bibfield  {author} {\bibinfo {author} {\bibfnamefont {W.-K.}\ \bibnamefont
  {Tse}}\ and\ \bibinfo {author} {\bibfnamefont {S.}~\bibnamefont
  {Das~Sarma}},\ }\href {\doibase 10.1103/PhysRevB.79.235406} {\bibfield
  {journal} {\bibinfo  {journal} {Phys. Rev. B}\ }\textbf {\bibinfo {volume}
  {79}},\ \bibinfo {pages} {235406} (\bibinfo {year} {2009})}\BibitemShut
  {NoStop}%
\bibitem [{\citenamefont {Borysenko}\ \emph {et~al.}(2010)\citenamefont
  {Borysenko}, \citenamefont {Mullen}, \citenamefont {Barry}, \citenamefont
  {Paul}, \citenamefont {Semenov}, \citenamefont {Zavada}, \citenamefont
  {Nardelli},\ and\ \citenamefont {Kim}}]{PhysRevB.81.121412}%
  \BibitemOpen
  \bibfield  {author} {\bibinfo {author} {\bibfnamefont {K.~M.}\ \bibnamefont
  {Borysenko}}, \bibinfo {author} {\bibfnamefont {J.~T.}\ \bibnamefont
  {Mullen}}, \bibinfo {author} {\bibfnamefont {E.~A.}\ \bibnamefont {Barry}},
  \bibinfo {author} {\bibfnamefont {S.}~\bibnamefont {Paul}}, \bibinfo {author}
  {\bibfnamefont {Y.~G.}\ \bibnamefont {Semenov}}, \bibinfo {author}
  {\bibfnamefont {J.~M.}\ \bibnamefont {Zavada}}, \bibinfo {author}
  {\bibfnamefont {M.~B.}\ \bibnamefont {Nardelli}}, \ and\ \bibinfo {author}
  {\bibfnamefont {K.~W.}\ \bibnamefont {Kim}},\ }\href {\doibase
  10.1103/PhysRevB.81.121412} {\bibfield  {journal} {\bibinfo  {journal} {Phys.
  Rev. B}\ }\textbf {\bibinfo {volume} {81}},\ \bibinfo {pages} {121412}
  (\bibinfo {year} {2010})}\BibitemShut {NoStop}%
\bibitem [{\citenamefont {Suzuura}\ and\ \citenamefont
  {Ando}(2002)}]{PhysRevB.65.235412}%
  \BibitemOpen
  \bibfield  {author} {\bibinfo {author} {\bibfnamefont {H.}~\bibnamefont
  {Suzuura}}\ and\ \bibinfo {author} {\bibfnamefont {T.}~\bibnamefont {Ando}},\
  }\href {\doibase 10.1103/PhysRevB.65.235412} {\bibfield  {journal} {\bibinfo
  {journal} {Phys. Rev. B}\ }\textbf {\bibinfo {volume} {65}},\ \bibinfo
  {pages} {235412} (\bibinfo {year} {2002})}\BibitemShut {NoStop}%
\bibitem [{\citenamefont {Jiang}\ \emph {et~al.}(2005)\citenamefont {Jiang},
  \citenamefont {Saito}, \citenamefont {Samsonidze}, \citenamefont {Chou},
  \citenamefont {Jorio}, \citenamefont {Dresselhaus},\ and\ \citenamefont
  {Dresselhaus}}]{PhysRevB.72.235408}%
  \BibitemOpen
  \bibfield  {author} {\bibinfo {author} {\bibfnamefont {J.}~\bibnamefont
  {Jiang}}, \bibinfo {author} {\bibfnamefont {R.}~\bibnamefont {Saito}},
  \bibinfo {author} {\bibfnamefont {G.~G.}\ \bibnamefont {Samsonidze}},
  \bibinfo {author} {\bibfnamefont {S.~G.}\ \bibnamefont {Chou}}, \bibinfo
  {author} {\bibfnamefont {A.}~\bibnamefont {Jorio}}, \bibinfo {author}
  {\bibfnamefont {G.}~\bibnamefont {Dresselhaus}}, \ and\ \bibinfo {author}
  {\bibfnamefont {M.~S.}\ \bibnamefont {Dresselhaus}},\ }\href {\doibase
  10.1103/PhysRevB.72.235408} {\bibfield  {journal} {\bibinfo  {journal} {Phys.
  Rev. B}\ }\textbf {\bibinfo {volume} {72}},\ \bibinfo {pages} {235408}
  (\bibinfo {year} {2005})}\BibitemShut {NoStop}%
\bibitem [{\citenamefont {Eidelman}\ and\ \citenamefont {Vul'}(2007)}]{EidVul}%
  \BibitemOpen
  \bibfield  {author} {\bibinfo {author} {\bibfnamefont {E.~D.}\ \bibnamefont
  {Eidelman}}\ and\ \bibinfo {author} {\bibfnamefont {A.~Y.}\ \bibnamefont
  {Vul'}},\ }\href {http://stacks.iop.org/0953-8984/19/i=26/a=266210}
  {\bibfield  {journal} {\bibinfo  {journal} {Journal of Physics: Condensed
  Matter}\ }\textbf {\bibinfo {volume} {19}},\ \bibinfo {pages} {266210}
  (\bibinfo {year} {2007})}\BibitemShut {NoStop}%
\bibitem [{\citenamefont {Ayache}\ \emph {et~al.}(1980)\citenamefont {Ayache},
  \citenamefont {de~Combarieu},\ and\ \citenamefont
  {Jay-Gerin}}]{PhysRevB.21.2462}%
  \BibitemOpen
  \bibfield  {author} {\bibinfo {author} {\bibfnamefont {C.}~\bibnamefont
  {Ayache}}, \bibinfo {author} {\bibfnamefont {A.}~\bibnamefont
  {de~Combarieu}}, \ and\ \bibinfo {author} {\bibfnamefont {J.~P.}\
  \bibnamefont {Jay-Gerin}},\ }\href {\doibase 10.1103/PhysRevB.21.2462}
  {\bibfield  {journal} {\bibinfo  {journal} {Phys. Rev. B}\ }\textbf {\bibinfo
  {volume} {21}},\ \bibinfo {pages} {2462} (\bibinfo {year}
  {1980})}\BibitemShut {NoStop}%
\bibitem [{\citenamefont {Sugihara}(1983)}]{PhysRevB.28.2157}%
  \BibitemOpen
  \bibfield  {author} {\bibinfo {author} {\bibfnamefont {K.}~\bibnamefont
  {Sugihara}},\ }\href {\doibase 10.1103/PhysRevB.28.2157} {\bibfield
  {journal} {\bibinfo  {journal} {Phys. Rev. B}\ }\textbf {\bibinfo {volume}
  {28}},\ \bibinfo {pages} {2157} (\bibinfo {year} {1983})}\BibitemShut
  {NoStop}%
\bibitem [{\citenamefont {Sugihara}\ \emph {et~al.}(1986)\citenamefont
  {Sugihara}, \citenamefont {Hishiyama},\ and\ \citenamefont
  {Ono}}]{PhysRevB.34.4298}%
  \BibitemOpen
  \bibfield  {author} {\bibinfo {author} {\bibfnamefont {K.}~\bibnamefont
  {Sugihara}}, \bibinfo {author} {\bibfnamefont {Y.}~\bibnamefont {Hishiyama}},
  \ and\ \bibinfo {author} {\bibfnamefont {A.}~\bibnamefont {Ono}},\ }\href
  {\doibase 10.1103/PhysRevB.34.4298} {\bibfield  {journal} {\bibinfo
  {journal} {Phys. Rev. B}\ }\textbf {\bibinfo {volume} {34}},\ \bibinfo
  {pages} {4298} (\bibinfo {year} {1986})}\BibitemShut {NoStop}%
\bibitem [{\citenamefont {Scarola}\ and\ \citenamefont
  {Mahan}(2002)}]{PhysRevB.66.205405}%
  \BibitemOpen
  \bibfield  {author} {\bibinfo {author} {\bibfnamefont {V.~W.}\ \bibnamefont
  {Scarola}}\ and\ \bibinfo {author} {\bibfnamefont {G.~D.}\ \bibnamefont
  {Mahan}},\ }\href {\doibase 10.1103/PhysRevB.66.205405} {\bibfield  {journal}
  {\bibinfo  {journal} {Phys. Rev. B}\ }\textbf {\bibinfo {volume} {66}},\
  \bibinfo {pages} {205405} (\bibinfo {year} {2002})}\BibitemShut {NoStop}%
\bibitem [{\citenamefont {Gurevich}(1946)}]{Gurevich46}%
  \BibitemOpen
  \bibfield  {author} {\bibinfo {author} {\bibfnamefont {L.}~\bibnamefont
  {Gurevich}},\ }\href@noop {} {\bibfield  {journal} {\bibinfo  {journal} {J.
  Phys. (USSR)}\ }\textbf {\bibinfo {volume} {9}},\ \bibinfo {pages} {4}
  (\bibinfo {year} {1946})}\BibitemShut {NoStop}%
\bibitem [{\citenamefont {Giovannetti}\ \emph {et~al.}(2008)\citenamefont
  {Giovannetti}, \citenamefont {Khomyakov}, \citenamefont {Brocks},
  \citenamefont {Karpan}, \citenamefont {van~den Brink},\ and\ \citenamefont
  {Kelly}}]{PhysRevLett.101.026803}%
  \BibitemOpen
  \bibfield  {author} {\bibinfo {author} {\bibfnamefont {G.}~\bibnamefont
  {Giovannetti}}, \bibinfo {author} {\bibfnamefont {P.~A.}\ \bibnamefont
  {Khomyakov}}, \bibinfo {author} {\bibfnamefont {G.}~\bibnamefont {Brocks}},
  \bibinfo {author} {\bibfnamefont {V.~M.}\ \bibnamefont {Karpan}}, \bibinfo
  {author} {\bibfnamefont {J.}~\bibnamefont {van~den Brink}}, \ and\ \bibinfo
  {author} {\bibfnamefont {P.~J.}\ \bibnamefont {Kelly}},\ }\href {\doibase
  10.1103/PhysRevLett.101.026803} {\bibfield  {journal} {\bibinfo  {journal}
  {Phys. Rev. Lett.}\ }\textbf {\bibinfo {volume} {101}},\ \bibinfo {pages}
  {026803} (\bibinfo {year} {2008})}\BibitemShut {NoStop}%
\bibitem [{\citenamefont {Nihira}\ and\ \citenamefont
  {Iwata}(2003)}]{PhysRevB.68.134305}%
  \BibitemOpen
  \bibfield  {author} {\bibinfo {author} {\bibfnamefont {T.}~\bibnamefont
  {Nihira}}\ and\ \bibinfo {author} {\bibfnamefont {T.}~\bibnamefont {Iwata}},\
  }\href {\doibase 10.1103/PhysRevB.68.134305} {\bibfield  {journal} {\bibinfo
  {journal} {Phys. Rev. B}\ }\textbf {\bibinfo {volume} {68}},\ \bibinfo
  {pages} {134305} (\bibinfo {year} {2003})}\BibitemShut {NoStop}%
\bibitem [{\citenamefont {Ma\~nes}(2007)}]{PhysRevB.76.045430}%
  \BibitemOpen
  \bibfield  {author} {\bibinfo {author} {\bibfnamefont {J.~L.}\ \bibnamefont
  {Ma\~nes}},\ }\href {\doibase 10.1103/PhysRevB.76.045430} {\bibfield
  {journal} {\bibinfo  {journal} {Phys. Rev. B}\ }\textbf {\bibinfo {volume}
  {76}},\ \bibinfo {pages} {045430} (\bibinfo {year} {2007})}\BibitemShut
  {NoStop}%
\bibitem [{\citenamefont {Anselm}(1982)}]{Anselm}%
  \BibitemOpen
  \bibfield  {author} {\bibinfo {author} {\bibfnamefont {A.~I.}\ \bibnamefont
  {Anselm}},\ }\href@noop {} {\emph {\bibinfo {title} {Introduction to
  Semiconductor Theory}}}\ (\bibinfo  {publisher} {Prentice Hall},\ \bibinfo
  {address} {Upper Saddle River, NJ},\ \bibinfo {year} {1982})\BibitemShut
  {NoStop}%
\bibitem [{\citenamefont {Lifshitz}\ and\ \citenamefont
  {Pitaevskii}(1981)}]{Dau10en}%
  \BibitemOpen
  \bibfield  {author} {\bibinfo {author} {\bibfnamefont {E.~M.}\ \bibnamefont
  {Lifshitz}}\ and\ \bibinfo {author} {\bibfnamefont {L.~P.}\ \bibnamefont
  {Pitaevskii}},\ }\href
  {http://www.amazon.com/exec/obidos/redirect?tag=citeulike07-20\&path=ASIN/0750626356}
  {\emph {\bibinfo {title} {{Course of Theoretical Physics, Volume X: Physical
  Kinetics}}}}\ (\bibinfo  {publisher} {Butterworth-Heinemann},\ \bibinfo
  {address} {Oxford},\ \bibinfo {year} {1981})\BibitemShut {NoStop}%
\bibitem [{\citenamefont {Chen}\ \emph {et~al.}(2008)\citenamefont {Chen},
  \citenamefont {Jang}, \citenamefont {Xiao}, \citenamefont {Ishigami},\ and\
  \citenamefont {Fuhrer}}]{SiO2Limit}%
  \BibitemOpen
  \bibfield  {author} {\bibinfo {author} {\bibfnamefont {J.-H.}\ \bibnamefont
  {Chen}}, \bibinfo {author} {\bibfnamefont {C.}~\bibnamefont {Jang}}, \bibinfo
  {author} {\bibfnamefont {S.}~\bibnamefont {Xiao}}, \bibinfo {author}
  {\bibfnamefont {M.}~\bibnamefont {Ishigami}}, \ and\ \bibinfo {author}
  {\bibfnamefont {M.~S.}\ \bibnamefont {Fuhrer}},\ }\href {\doibase
  http://dx.doi.org/10.1038/nnano.2008.58} {\bibfield  {journal} {\bibinfo
  {journal} {Nat. Nano}\ }\textbf {\bibinfo {volume} {3}},\ \bibinfo {pages}
  {206} (\bibinfo {year} {2008})}\BibitemShut {NoStop}%
\bibitem [{\citenamefont {Qiu}\ and\ \citenamefont {Ruan}(2012)}]{MDphonons}%
  \BibitemOpen
  \bibfield  {author} {\bibinfo {author} {\bibfnamefont {B.}~\bibnamefont
  {Qiu}}\ and\ \bibinfo {author} {\bibfnamefont {X.}~\bibnamefont {Ruan}},\
  }\href {\doibase 10.1063/1.4712041} {\bibfield  {journal} {\bibinfo
  {journal} {Applied Physics Letters}\ }\textbf {\bibinfo {volume} {100}},\
  \bibinfo {eid} {193101} (\bibinfo {year} {2012})}\BibitemShut {NoStop}%
\bibitem [{\citenamefont {Fan}\ \emph {et~al.}(2012)\citenamefont {Fan},
  \citenamefont {Zheng}, \citenamefont {Chihaia}, \citenamefont {Shen},\ and\
  \citenamefont {Kuo}}]{GrapheneSiO2}%
  \BibitemOpen
  \bibfield  {author} {\bibinfo {author} {\bibfnamefont {X.~F.}\ \bibnamefont
  {Fan}}, \bibinfo {author} {\bibfnamefont {W.~T.}\ \bibnamefont {Zheng}},
  \bibinfo {author} {\bibfnamefont {V.}~\bibnamefont {Chihaia}}, \bibinfo
  {author} {\bibfnamefont {Z.~X.}\ \bibnamefont {Shen}}, \ and\ \bibinfo
  {author} {\bibfnamefont {J.-L.}\ \bibnamefont {Kuo}},\ }\href
  {http://stacks.iop.org/0953-8984/24/i=30/a=305004} {\bibfield  {journal}
  {\bibinfo  {journal} {Journal of Physics: Condensed Matter}\ }\textbf
  {\bibinfo {volume} {24}},\ \bibinfo {pages} {305004} (\bibinfo {year}
  {2012})}\BibitemShut {NoStop}%
\bibitem [{\citenamefont {Brako}\ \emph {et~al.}(2010)\citenamefont {Brako},
  \citenamefont {Å $\check{S}$ok$\check{c}$evi$\acute{c}$}, \citenamefont
  {Lazi$\acute{c}$},\ and\ \citenamefont {Atodiresei}}]{GrapheneIr111}%
  \BibitemOpen
  \bibfield  {author} {\bibinfo {author} {\bibfnamefont {R.}~\bibnamefont
  {Brako}}, \bibinfo {author} {\bibfnamefont {D.}~\bibnamefont
  {Å $\check{S}$ok$\check{c}$evi$\acute{c}$}}, \bibinfo {author} {\bibfnamefont
  {P.}~\bibnamefont {Lazi$\acute{c}$}}, \ and\ \bibinfo {author} {\bibfnamefont
  {N.}~\bibnamefont {Atodiresei}},\ }\href
  {http://stacks.iop.org/1367-2630/12/i=11/a=113016} {\bibfield  {journal}
  {\bibinfo  {journal} {New Journal of Physics}\ }\textbf {\bibinfo {volume}
  {12}},\ \bibinfo {pages} {113016} (\bibinfo {year} {2010})}\BibitemShut
  {NoStop}%
\bibitem [{\citenamefont {Wang}\ \emph {et~al.}(2008)\citenamefont {Wang},
  \citenamefont {Ni}, \citenamefont {Yu}, \citenamefont {Shen}, \citenamefont
  {Wang}, \citenamefont {Wu}, \citenamefont {Chen},\ and\ \citenamefont
  {Shen~Wee}}]{doi:10.1021/jp8008404}%
  \BibitemOpen
  \bibfield  {author} {\bibinfo {author} {\bibfnamefont {Y.~y.}\ \bibnamefont
  {Wang}}, \bibinfo {author} {\bibfnamefont {Z.~h.}\ \bibnamefont {Ni}},
  \bibinfo {author} {\bibfnamefont {T.}~\bibnamefont {Yu}}, \bibinfo {author}
  {\bibfnamefont {Z.~X.}\ \bibnamefont {Shen}}, \bibinfo {author}
  {\bibfnamefont {H.~m.}\ \bibnamefont {Wang}}, \bibinfo {author}
  {\bibfnamefont {Y.~h.}\ \bibnamefont {Wu}}, \bibinfo {author} {\bibfnamefont
  {W.}~\bibnamefont {Chen}}, \ and\ \bibinfo {author} {\bibfnamefont {A.~T.}\
  \bibnamefont {Shen~Wee}},\ }\href {\doibase 10.1021/jp8008404} {\bibfield
  {journal} {\bibinfo  {journal} {The Journal of Physical Chemistry C}\
  }\textbf {\bibinfo {volume} {112}},\ \bibinfo {pages} {10637} (\bibinfo
  {year} {2008})},\ \Eprint
  {http://arxiv.org/abs/http://pubs.acs.org/doi/pdf/10.1021/jp8008404}
  {http://pubs.acs.org/doi/pdf/10.1021/jp8008404} \BibitemShut {NoStop}%
\bibitem [{\citenamefont {Dideykin}\ \emph {et~al.}(2011)\citenamefont
  {Dideykin}, \citenamefont {Aleksenskiy}, \citenamefont {Kirilenko},
  \citenamefont {Brunkov}, \citenamefont {Goncharov}, \citenamefont
  {Baidakova}, \citenamefont {Sakseev},\ and\ \citenamefont
  {Ya.Vul'}}]{Dideykin2011105}%
  \BibitemOpen
  \bibfield  {author} {\bibinfo {author} {\bibfnamefont {A.}~\bibnamefont
  {Dideykin}}, \bibinfo {author} {\bibfnamefont {A.}~\bibnamefont
  {Aleksenskiy}}, \bibinfo {author} {\bibfnamefont {D.}~\bibnamefont
  {Kirilenko}}, \bibinfo {author} {\bibfnamefont {P.}~\bibnamefont {Brunkov}},
  \bibinfo {author} {\bibfnamefont {V.}~\bibnamefont {Goncharov}}, \bibinfo
  {author} {\bibfnamefont {M.}~\bibnamefont {Baidakova}}, \bibinfo {author}
  {\bibfnamefont {D.}~\bibnamefont {Sakseev}}, \ and\ \bibinfo {author}
  {\bibfnamefont {A.}~\bibnamefont {Ya.Vul'}},\ }\href {\doibase
  10.1016/j.diamond.2010.10.007} {\bibfield  {journal} {\bibinfo  {journal}
  {Diamond and Related Materials}\ }\textbf {\bibinfo {volume} {20}},\ \bibinfo
  {pages} {105 } (\bibinfo {year} {2011})}\BibitemShut {NoStop}%
\bibitem [{\citenamefont {Liang}\ \emph {et~al.}(2007)\citenamefont {Liang},
  \citenamefont {Fu},\ and\ \citenamefont {Chou}}]{doi:10.1021/nl072566s}%
  \BibitemOpen
  \bibfield  {author} {\bibinfo {author} {\bibfnamefont {X.}~\bibnamefont
  {Liang}}, \bibinfo {author} {\bibfnamefont {Z.}~\bibnamefont {Fu}}, \ and\
  \bibinfo {author} {\bibfnamefont {S.~Y.}\ \bibnamefont {Chou}},\ }\href
  {\doibase 10.1021/nl072566s} {\bibfield  {journal} {\bibinfo  {journal} {Nano
  Letters}\ }\textbf {\bibinfo {volume} {7}},\ \bibinfo {pages} {3840}
  (\bibinfo {year} {2007})},\ \Eprint
  {http://arxiv.org/abs/http://pubs.acs.org/doi/pdf/10.1021/nl072566s}
  {http://pubs.acs.org/doi/pdf/10.1021/nl072566s} \BibitemShut {NoStop}%
\end{thebibliography}
%

\end{document}